\documentclass{article}
\usepackage{spconf,amsmath,graphicx, booktabs, multirow, caption,xfrac}
\usepackage[inline]{enumitem}
\usepackage{stfloats}

\usepackage{hyperref}
\usepackage{cleveref}
\newcommand\inv[1]{#1\raisebox{1.15ex}{$\scriptscriptstyle-\!1$}}

\title{Duration robust weakly supervised sound event detection}
\name{Heinrich Dinkel and Kai Yu\thanks{This work has been supported by the Major Program of National Social Science Foundation of China (No.18ZDA293). Experiments have been carried out on the PI supercomputer at Shanghai Jiao Tong University.}}
\address{MoE Key Lab of Artificial Intelligence\\
	SpeechLab, Department of Computer Science and Engineering\\
	Shanghai Jiao Tong University, Shanghai, China\\
	\texttt{\small{heinrich.dinkel@gmail.com, kai.yu@sjtu.edu.cn}}
    }

\begin{document}
\topmargin=0mm
%
\maketitle

\begin{abstract}
Task 4 of the DCASE2018 challenge demonstrated that substantially more research is needed for a real-world application of sound event detection. 
Analyzing the challenge results it can be seen that most successful models are biased towards predicting long (e.g., over 5s) clips.
This work aims to investigate the performance impact of fixed-sized window median filter post-processing and advocate the use of double thresholding as a more robust and predictable post-processing method. 
Further, four different temporal subsampling methods within the CRNN framework are proposed: mean-max, $\alpha$-mean-max, $L^p$-norm and convolutional. 
We show that for this task subsampling the temporal resolution by a neural network enhances the F1 score as well as its robustness towards short, sporadic sound events.
Our best single model achieves 30.1\% F1 on the evaluation set and the best fusion model $32.5\%$, while being robust to event length variations.
\end{abstract}
\noindent\textbf{Index Terms}: weakly supervised sound event detection, convolutional neural networks, recurrent neural networks, semi-supervised duration estimation

\section{Introduction}
\label{sec:intro}
Sound event detection (SED) is concerned with the classification and localization of a particular audio event (e.g., dog barking, alarm ringing) such that each event is assigned an onset (start), offset (end) and label (tagging). 
In particular, this paper focuses on weakly-supervised SED (WSSED), a semi-supervised task, which has only access to clip-based labels during training, yet needs to predict onsets and offsets during evaluation.

SED can be used for query-based sound retrieval \cite{Font2018}, smart cities and homes \cite{Bello2018, Krstulovic2018}. 
In contrast to similar tasks such as speech/speaker recognition, the recorded audio properties are vast, often noisy, can overlap and be assigned multiple events at once.
Recent interest in SED has risen due to challenges such as the Detection and Classification of
Acoustic Scenes and Events (DCASE) challenge. 
In this work, we focus on sound event detection within domestic environments, specifically task 4 of the DCASE2018 challenge\cite{Serizel2018}. 


Recently, much research attention has been brought up in order to improve CRNN performance \cite{Cakr2017, Iqbal2018}. 
Kothinti \cite{Kothinti2018} presented an interesting approach by separating SED into an unsupervised onset and offset estimation problem using conditional restricted Boltzmann machines (c-RBM) along with a supervised label prediction using CRNN. 
The results of 30\% F1 development and 25 \% F1 evaluation performance on the DCASE2018 task4 dataset indicate the robustness of this approach.
Wang \cite{Wang2018a} modified connectionist temporal classification (CTC)\cite{DBLP:conf/icml/GravesFGS06} in order to enable capturing long and short events effectively.
An essential piece of work done in \cite{lin2019specialized}, analyzed the usage of attention-based high-level feature disentangling.
Other studies in \cite{Cances_Post_processing} analyzed the impact of several post-processing methods, specifically the estimation of an event-dependent threshold, in regards to WSSED.
Their results indicate that the choice of post-processing is crucial, which we also testify in this work, by an increase of absolute 10.9\% in F1 score on the DCASE2018 task 4 dataset.
Lastly, \cite{Pellegrini2019} introduced a cosine penalty between different time-event predictions of the CRNN aiming to enhance the per time step discriminability of each event.
This idea is similar to large margin softmax (L-softmax)\cite{liu2016large} and resulted in an F1 score of $32.42\%$ on the DCASE2018 task4 development dataset.

In contrast to previous work, we aim to enhance WSSED performance, specifically regarding short-duration events, by estimating event duration in a robust manner.
This paper is organized as follows: In \Cref{sec:contribution} we state current problems for state-of-the-art event-based SED and propose our idea to alleviate these problems.
Further, in \Cref{sec:Experiments} the experimental setup and experiments are run, culminating in \Cref{sec:results}, where results are shown and analyzed.

\section{Contribution}
\label{sec:contribution}

In our point of view, WSSED`s main difficulty is the incapability of estimating appropriate event duration, due to the lack of any prior event knowledge.
Specifically, the estimation of short, sporadic events such as dog barking is more difficult than estimating long, continuous events.
In this work, we identify and analyze three key problems within WSSED: 1) During training weak label estimates are obtained via mean-pooling the temporal dimension.
This approach benefits long events and neglects short ones\cite{Wang2018};
2) During inference, per-frame predictions are post-processed to smooth event predictions, using median filtering, which is shown to benefit long events further;
3) Neural network predictions are made on a fine scale (e.g., 20ms). Due to the noisy nature of this task, post-processing is necessary in order to obtain coherent predictions. However, post-processing cannot be learned by the network directly. 

In this work we aim to alleviate the problems stated above by 1) incorporating linear softmax \cite{Wang2018} as the default temporal pooling method 2) Using double threshold as a window-independent filtering alternative to median filtering 3) Subsampling the temporal resolution of our neural network predictions in order to learn event boundaries directly.
Furthermore, we show that the previous best model submitted to the DCASE2018 task 4 challenge \cite{Lu2018} is biased towards long event predictions and propose a duration robust alternative.

\subsection{Metric}
\label{sub: metrics}

This work solely uses event-based F1 score \cite{Mesaros2016_MDPI} as the primary evaluation metric, which requires predictions to be smooth (contiguous) and penalizes irregular or disjoint predictions. 
In order to loosen the strictness of this measure, a flexible onset (t-collar) of 200ms, as well as an offset of at most 20\% of the events` duration, is considered as valid.

\section{Experiments}%
\label{sec:Experiments}

Regarding feature extraction, $64$-dimensional log-mel spectrograms (LMS) were used in this work. Each LMS sample was extracted by a $2048$ point Fourier transform every $20$ms with a window size of $40$ms.
Since our network does process the input in a sequential fashion (e.g., convolutions over spatial and frequency dimensions), padding needs to be applied to ensure a fixed input size. 
Batch-wise zero padding is applied, which essentially is equal to pad the entire data to the maximal length of $499$ frames (10 s). 
Since the Speech class accounts for nearly 25\% of the total training set, random oversampling is utilized by assigning each class a weight inverse proportional to its occurrence count. 
Since during training, hard labels are unknown, a final temporal pooling function needs to be utilized in order to reduce an input clip`s temporal dimension to a single vector representing class probabilities.
Work in \cite{Wang2018} proposed linear softmax (LS), a parameter-free and effective temporal pooling function for WSSED.

\begin{equation}\label{eq:linear_softmax}
	y(c) = \frac{\sum_{t}^T y_t(c)^2}{\sum_{t}^T y_t(c)}
\end{equation}

LS is defined as in \Cref{eq:linear_softmax}, where $y_t(c) \in [0,1]$ is the output probability of event-class $c$ at timestep $t$. 
Since LS is only dependent on the per-frame probability and not number of frames, it is more robust to length variations than traditional mean pooling.
Standard binary cross entropy (BCE) is used as the training criterion.
Our model follows a CRNN approach and can be seen in \Cref{tab:arch}, where BiGRU represents a bidirectional GRU recurrent neural network.
LS is only used during training, while during inference post-processing methods are applied (see \Cref{sub:postprocessing}).
For each subsampling layer $(s_1,s_2,s_3, s_4)$, we employ a different time subsampling factor, notated as $P: \big\{ \left( s_1, s_2, s_3, s_4 \right) \in \big\{ 1,2 \big\}^4: s_i \geq s_j, 1 \leq i \leq j \leq 4 \big\}, \left( s_1, s_2, s_3, s_4 \right) \mapsto \prod_{i=1}^4 s_i $.
Moreover, the inverse $\inv{P}$, maps a subsampling factor to a subsampling layer sequence.
Here we only subsample at most by a factor of $2$ at each layer, while the feature dimension $D$ is always halved. 
Five different subsampling configurations are introduced by $\mathcal{S}_k = \inv{P}(k), k = 1,2,4,8,16$, where $\mathcal{S}_1$ represents no temporal subsampling and $\mathcal{S}_{16}$ represents subsampling by a factor of $16$.

\begin{table}[ht]
	\renewcommand*{\arraystretch}{1.05}
	\centering
	\begin{tabular}{l|r}
		Layer & Parameter\\
		\hline\hline
		Block1 & $16$ Channel, $3\times3$ Kernel \\
		Subsample1 & $s_1 \downarrow 2$ \\
		\hline
        Block2 & $32$ Channel, $3\times3$ Kernel \\
		Subsample2 & $s_2 \downarrow 2$ \\
		\hline

        Block3 & $128$ Channel, $3\times3$ Kernel\\
        Subsample3 & $s_3 \downarrow 2$ \\
		\hline

		Block4 & $128$ Channel, $3\times3$ Kernel \\
		Subsample4 & $s_4 \downarrow 2$ \\
		\hline

        Block5 & $128$ Channel, $3\times3$ Kernel \\
		Dropout & $30\%$ \\
		\hline

		BiGRU & $128$ Units  \\
		Linear & $10$ Units  \\
		LS & $ $ 

	\end{tabular}
	\caption{CRNN architecture used in this work. One block refers to an initial batch normalization, then a convolution and lastly a ReLU activation. All convolutions use padding in order to preserve the input size. The notation $s_k \downarrow 2$ represents subsampling temporal dimension by $s_k$ as well as halving the feature dimension $D$.}
	\label{tab:arch}
\end{table}

Training was done using Adam optimization with an initial learning rate of 0.001. 
The learning rate was reduced if the model did not improve on the held-out set for 10 epochs. 
If the learning rate dropped below $1\mathrm{e}{-7}$ training was terminated. 
We used pytorch\cite{paszke2017automatic} as our deep neural network tool\footnote{Code is available at \href{www.github.com/richermans/dcase2018\_pooling}{github.com/richermans/dcase2018\_pooling}}.

\subsection{Dataset}%
\label{sub:Dataset}
Experiments are conducted on the DCASE2018 task 4 dataset, sampled from the larger AudioSet \cite{gemmeke2017audio}.
The dataset consists of a weakly-labelled training set, hard-labelled development and evaluation sets, as well as the unlabelled indomain and outdomain subsets. 
This paper only uses training, development and evaluation subsets.
The DCASE2018 task 4 dataset consists of 1578 training clips with 10 class labels, as well as 242 development clips situated in domestic environments.
Ten classes need to be estimated: Speech, Cat, Dog, Running water, Vacuum cleaner, Frying, Electric shaver toothbrush, Blender and Alarm bell.

\begin{figure}[ht]
    \centering
    \includegraphics[width = 0.48\textwidth]{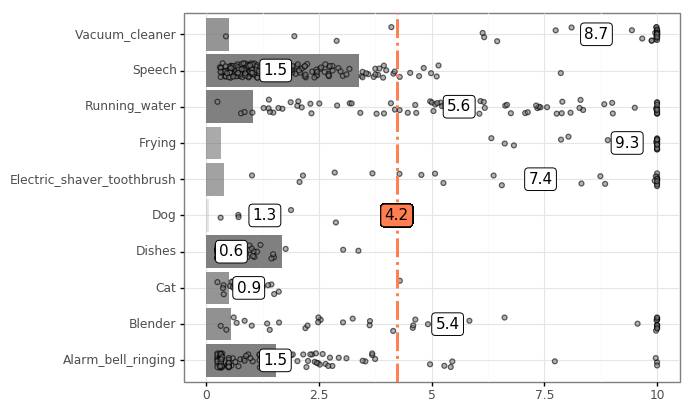}
    \caption{\small{DCASE18 Task4 development data length distribution. Average per-class duration (number in box) as well as the average data duration (colored box) are given. Each gray dot represents a single clip duration, and each bar relative class event occupation within the dataset.}}
    \label{fig: data_dist}
\end{figure}

Analyzing the development data distribution in \Cref{fig: data_dist} reveals that the dataset can be effectively split into long-duration events (Vacuum cleaner, Running water, Frying, Electric shaver toothbrush and Blender) and short duration events (Speech, Dog, Dishes, Cat and Alarm bell).

\subsection{Post-processing}
\label{sub:postprocessing}

During inference, post-processing is applied to smooth each event-class probability sequence.
Here we aim to investigate the effect of two post-processing methods regarding their hyperparameters: standard median filtering and the proposed double thresholding towards the model performance.
Two sets of experiments with mean and max subsampling were run using the default model configuration $\mathcal{S}_1$, meaning no temporal subsampling is applied.

\vspace{-0.3cm}
\subsubsection{Median filtering}%
\label{sub:baseline}

Standard median filtering first preprocesses $y_t(c)$ by a threshold $\phi, y_t(c) > \phi$. 
Then a median filter of size $\omega$ is applied in a rolling window fashion onto the sequence in order to smooth the frame predictions.
Here, the threshold value is set to $\phi = 0.5$ and two different median filter size configurations $\omega\in\{1,51\}$ were investigated.
A window size of 1 represents no filtering, while 51 is chosen as the default value.

\begin{table}[!ht]
  \centering
  \begin{tabular}{r|ccc|ccc}
     & \multicolumn{3}{|c|}{$\omega = 51$} & \multicolumn{3}{|c}{$\omega = 1$}\\
     \hline
      Sub & Short & Long & Avg & Short & Long & Avg \\
	\hline\hline
      Max & 26.18 & \textbf{23.78} & \textbf{24.98} & 26.4 & 11.94 & 19.17  \\
      Mean & 20.46 & 20.18 & 20.32 & \textbf{28.26} & 10.24 & 19.25
  \end{tabular}
  \caption{Development F1-scores for different window size $\omega$ values of a median filter with respect to long and short clips.}
  \label{tab:inital_results}
\end{table}

As our initial experiments (\Cref{tab:inital_results}) suggest, filtering with a window size of $\omega=51$ leads to an overall performance increase for both mean and max subsampling networks. 
However, this increase solely stems from a shift in focus from short clips to long ones. 
For mean subsampling, the short clip F1 score drops by 8\% absolute, while long clip performance increases by 10\% absolute. 
We, therefore, conclude that a larger $\omega$ correlates with an overall performance increase, at the cost of performance on short clips.
One of the major downsides of median filtering is that it can potentially erase model predictions, e.g., high-confidence model estimates, as well as shift previously predicted event-boundaries.
In this work we advocate the use of double threshold (see \Cref{sub:double_threshold}) in order to ameliorate the median filtering problems, that is the use a post-processing filter that does not erase high confidence estimates or shifts an event boundary.

\vspace{-0.43cm}
\subsubsection{Double threshold}%
\label{sub:double_threshold}
This technique uses two thresholds $\phi_{\text{low}}, \phi_{\text{hi}}$.
Double threshold first sweeps over an output probability sequence and marks all values being larger than $\phi_{\text{hi}}$ as being valid predictions.
Then it enlarges the marked frames by searching for all adjacent, continuous predictions being larger than $\phi_{\text{low}}$.
\begin{table}[!h]
    \centering
    \begin{tabular}{c|ccc|ccc}
     & \multicolumn{3}{|c|}{$\omega = 51$} & \multicolumn{3}{c}{$\omega = 1$}\\
     \hline
      Sub & Short & Long & Avg & Short & Long & Avg \\
        \hline\hline
      Max & 16.82 & \textbf{33.22} & 25.02 & 31.36 & 28.52 & 29.94\\
      Mean & 17.28 & 31.48 & 25.29 & \textbf{33.74} & 27.88 & \textbf{30.81}\\
    \end{tabular}
    \caption{Comparison of double thresholding with different window sizes ($\omega$) on the development set.}
    \label{tab:double_threshold}
\end{table}
Double threshold can also incorporate a window size $\omega$, but with a different purpose from standard median filtering. 
Here, a window size of $\omega$ represents the number of frames being connected after thresholding. 
In this paper we exclusively set $\phi_{\text{low}} = 0.2, \phi_{\text{hi}} = 0.75$.
As the results in \Cref{tab:double_threshold} indicate, double threshold provides an overall better and duration robust performance compared to fixed-sized median filtering, without being affected by a non-optimal $\omega$.
Due to these results, all following experiments are run with $\omega = 1$, effectively neglecting the influence of $\omega$.

Both mean and max subsampling seem to exhibit similar performance. Thus this work investigated the use of joint mean and max subsampling, as seen in \Cref{sec:subsampling}.

\begin{table*}[tp]
    \centering
\begin{tabular}{l|rr|rr|rr|rr|rr|rr}
Configuration $\mathcal{S}_k$ &     \multicolumn{2}{|c|}{1}  &     \multicolumn{2}{c|}{2}  &     \multicolumn{2}{c|}{4}  &     \multicolumn{2}{c|}{8}  &     \multicolumn{2}{c|}{16} & \multicolumn{2}{c}{Fusion}($2,4,8$), \\
\hline
Subset           &  dev & eval & dev & eval & dev & eval& dev & eval & dev & eval & dev & eval \\
\hline\hline
Winner-2018 & - & - & - & - &  & & - & - & - & - & 25.90 & 32.40 \\
\hline
Conv  &   27.26&14.97 &  23.04&19.95 &  32.05&22.46 &  24.80&21.13 &  16.39&17.07 & 25.26 & 23.68 \\
LP  &  28.82&23.29 &  32.30&27.46 &  35.34&\textbf{30.81} &  33.14&28.00 &  21.97&21.65 & 35.26 & 32.21 \\
MM &  30.35&24.72 &  35.64&29.80 &  27.98&25.14 &  31.15&28.20 &  20.11&21.83 & 36.29 & 31.01 \\
$\alpha$-MM &  23.22&20.13 &  \textbf{36.00}&27.93 &  32.92&30.72 &  31.76&27.54 &  24.39&23.00 & \textbf{36.44} & \textbf{32.52} \\
\end{tabular}
\caption{Results for all four proposed subsampling types. Fusion is done by averaging the model outputs of $k=2,4,8$. The Winner-2018 system is a fusion system. Results highlighted in bold are the best in class. }
    \label{tab:pooling_results}
\end{table*}

\subsection{Subsampling}
\label{sec:subsampling}

One contribution of this work is to investigate appropriate subsampling layers.

\begin{table}[ht]
	\centering
	\begin{tabular}{r | c}
		Name & Formulation \\
		\hline\hline
		MM & $\text{mean}(x) + \max(x)$ \\
		$\alpha$-MM & $\alpha \max(x) + \left(1-\alpha\right) \text{mean}(x)  $\\
		LP & $\sqrt[p]{x^{p}}$ \\
		Conv & $\mathbf{W}x $ \\
	\end{tabular}
    \caption{Proposed subsampling layers. $\alpha$ is learned\cite{lee2016generalizing}. $p$ in $L^p$-norm is empirically set to $4$.}
	\label{tab:pooling_layers}
\end{table}
Here we propose four subsampling schemes (\Cref{tab:pooling_layers}), where $x$ represents the region of the image being pooled.
The methods consist of averaged mean-max subsampling (MM) used in the LightCNN framework\cite{wu2018light}, the convex combination of MM and $\alpha$-mean-max denoted as ($\alpha$-MM)\cite{lee2016generalizing}.
$L^p$-norm subsampling can be seen as a generalization of mean and max subsampling, where $p=1$ equals to mean and $p=\infty$ equals to max subsampling. 
Lastly, we investigate a weighted subsampling method implemented as strided convolutions with kernel size $K\times K$ and stride $K\times K$.

\section{Experimental Results}%
\label{sec:results}

In this work, we compare our proposed approach to the winner of the DCASE2018 task 4 challenge \cite{Lu2018} on the development and evaluation set results. 
The winner system differs from our approach in two major ways:
\begin{enumerate*}[label=\itshape\arabic*\upshape)]
    \item Uses \textit{semi-supervised} median filtering, where the window length $\omega$ is estimated according to the labelled development data.
    \item Uses the additionally provided \textit{unlabelled} in-domain dataset for model training.
\end{enumerate*}
Results in \Cref{tab:pooling_results} indicate that subsampling improves F1-score performance consistently up until $\mathcal{S}_4$ for all subsampling methods except MM, which stops at $\mathcal{S}_8$. The largest tested factor of $\mathcal{S}_{16}$ shows no improvement for all proposed methods, due to the temporal resolution reaching 320ms, which is longer than the t-collar of 200ms, leading to possible event label skips.
More importantly, the majority of our models produce stable scores between development and evaluation set and only lose around 5\% in absolute between the datasets. 

\begin{table}[ht]
    \centering
    \begin{tabular}{c|ccc|c}
        Type & Short & Long & Gap & Avg \\
         \hline
         \hline
        2018-Winner& 23.32 & \textbf{40.36} & 17.04 & 32.40  \\
         \hline
        Conv & 14.80 & 32.50 & 17.70 & 23.68 \\
        LP&  \textbf{30.20} & 34.22 & \textbf{4.02} & 32.21 \\
        MM & 27.92 & 34.14 & 6.22 & 31.03 \\
        $\alpha$-MM& 29.66 & 35.40 & 5.74 & \textbf{32.52}
    \end{tabular}
    \caption{Short and Long clip results for evaluation data. Gap is the absolute difference between long and short clip F1 scores. All shown results are model fusions.}
    \label{tab:fusion_results}
\end{table}

The results proposed in \Cref{tab:pooling_results} seem to indicate that strided convolution is largely outperformed by traditional subsampling methods, which we believe is due to the additional parameters to the limited amount of data provided in this challenge. 
As can be seen in \Cref{tab:fusion_results}, the previously best performing system is biased towards predicting long clips having a gap of $17\%$ absolute between short and long ones. 
Our proposed systems can reduce the short-long clip gap to as low as $4.02\%$, while performing equally as well as the 2018-Winner system.
In future research, we would like to investigate dynamic subsampling strategies further.

\section{Conclusions}

This paper shows that current WSSED focus on long term events in order to enhance performance while neglecting short events. 
It is shown that a bias towards long events for the DCASE2018 Task4 winner system exists.
Three potential reasons for this bias are tracked down: 1) Temporal mean pooling 2) Fixed sized median filtering 3) Training neural networks on a fine-grained scale. 
We alleviate each particular problem by advocating the use of 1) Linear softmax pooling 2) Double threshold filtering 3) Temporal subsampling to a lower scale.
Double thresholding as a post-processing method is shown to broadly outperform median filtering in terms of both robustness to duration and overall performance while being unaffected by choice of window size.
The standard CRNN model is modified to subsample the temporal resolution up until a factor of $16$ and shown to improve performance up until a factor of $8$.
Further, variations of mean and max subsampling are shown to enhance the performance on development and evaluation sets.
Our best single model achieves 30.8\%, while the best model fusion obtains 32.5\% F1.
Furthermore, our proposed method reduces the gap between short and long clip to 4\% F1.

\bibliographystyle{IEEEbib}
\bibliography{strings,refs}

\end{document}